\documentclass{article}

\usepackage[english]{babel}

\usepackage[letterpaper,top=2cm,bottom=2cm,left=3cm,right=3cm,marginparwidth=1.75cm]{geometry}

\usepackage{amsmath}
\usepackage{graphicx}
\usepackage[colorlinks=true, allcolors=blue]{hyperref}

\usepackage{bm}
\usepackage{xcolor}
\usepackage{comment}
\usepackage{hyperref}
\usepackage{caption}
\usepackage{amsmath}
\usepackage{bbold}
\usepackage{etoolbox}
\usepackage{adjustbox}
\usepackage{makecell}
\usepackage{subcaption}
\usepackage{textcomp}
\usepackage[symbol]{footmisc}
\usepackage{booktabs}
\usepackage{multirow}
\usepackage{algorithm}
\usepackage{algorithmic}
 \usepackage{setspace}
 \usepackage{textcomp}
\usepackage[sort&compress,numbers]{natbib}
\providecommand{\keywords}[1]{\textbf{Keywords:} #1}

\title{A Layered Swarm Optimization Method for Fitting Battery Thermal Runaway Models to Accelerating Rate Calorimetry Data}

\author{\stepcounter{footnote}Saakaar Bhatnagar$^{a}$\thanks{Corresponding Author} \\ {\small \textit{sbhatnagar@altair.com}} \and Andrew Comerford$^{a}$ \\ { \small \textit{acomerford@altair.com}} \and Zelu Xu$^{a}$ \\ { \small \textit{zxu@altair.com}}  \and Simone Reitano$^{b}$ \\ { \small \textit{simone.reitano@beond.net}} \and Luigi Scrimieri$^{b}$ \\ { \small \textit{luigi.scrimieri@beond.net}} \and Luca Giuliano$^{b}$ \\ { \small \textit{luca.giuliano@beond.net}} \and Araz Banaeizadeh$^{a}$ \\ { \small \textit{araz@altair.com}}  \\ }

\date{$^{a}${\small Altair Engineering Inc., 640 W. California Ave, Sunnyvale, CA, USA}\\ $^{b}${\small BeonD Srl - Corso Castelfidardo, 30A, 10129 Torino, Italy}} %leave blank

\begin{document}

\maketitle

\keywords{ Thermal Runaway, Population Methods, Model Fitting, Artificial Intelligence, Battery Safety } \\

\begin{abstract} \centering Thermal runaway in lithium-ion batteries is a critical safety concern for the battery industry due to its potential to cause uncontrolled temperature rises and subsequent fires that can engulf the battery pack and its surroundings. Modeling and simulation offer cost-effective tools for designing strategies to mitigate thermal runaway. Accurately simulating the chemical kinetics of thermal runaway, commonly represented by systems of Arrhenius-based Ordinary Differential Equations (ODEs), requires fitting kinetic parameters to experimental calorimetry data, such as Accelerating Rate Calorimetry (ARC) measurements. However, existing fitting methods often rely on empirical assumptions and simplifications that compromise generality or require manual tuning during the fitting process. Particle Swarm Optimization (PSO) offers a promising approach for directly fitting kinetic parameters to experimental data. Yet, for systems created by multiple Arrhenius ODEs, the computational cost of fitting using a brute-force approach that searches the entire parameter space simultaneously can become prohibitive. This work introduces a divide-and-conquer approach based on PSO to fit N-equation Arrhenius ODE models to ARC data. The proposed method achieves more accurate parameter fitting compared to the brute-force method while maintaining low computational costs. The method is analyzed using two distinct ARC datasets, and the resulting models are further validated through simulations of 3D ARC and oven tests, showing excellent agreement with experimental data and alignment with expected trends.
\end{abstract}

\section{Introduction}
\label{sec:intro}

Thermal runaway in battery packs is a significant safety concern, particularly in high-energy applications such as electric vehicles (EVs). This phenomenon arises under thermal abuse conditions, triggering exothermic degradation reactions in battery components, including anode decomposition, cathode conversion, SEI decomposition, and electrolyte breakdown \cite{feng2018thermal,spotnitz2003abuse}. Common causes include physical damage, internal short circuits, overcharging, or overheating due to extreme temperature exposure. The heat generated under these conditions can initiate a chain reaction where neighbouring cells enter a self-heating state, leading to thermal runaway. This propagation can affect an entire battery module or pack, posing serious safety risks. These challenges are becoming increasingly critical as the industry transitions to higher power and energy-density cells \cite{Golubkov2014,feng2018thermal}. To mitigate such risks, cell and pack manufacturers must implement stringent safety protocols, and simulation-driven design plays a pivotal role in optimizing battery designs, enabling efficient thermal analysis of innovative heat shield materials to evaluate their effectiveness in preventing thermal runaway propagation.\\

Typically, thermal runaway modeling represents the exothermic decomposition reactions as a system of Arrhenius Ordinary Differential Equations (ODEs)\cite{Hatchard2001,spotnitz2003abuse,Kim2007}. The reaction rates, for each decomposition reaction, combined with the reaction enthalpy, enable the prediction of chemical heat generation under abuse conditions. For each Arrhenius ODE, the kinetic parameters and heat of reaction must be determined. A popular and widely cited model in literature is the one proposed by \citet{Hatchard2001}. This model used calorimetry-derived data (from ARC and DSC) to model thermal runaway in an 18650 Lithium Cobalt Oxide (LCO) cylindrical cell, incorporating decomposition reactions for the anode, cathode, and SEI . The model treated the jellyroll of the cell as a single lumped temperature and showed excellent agreement with experimental data from an oven test. This model laid the foundation for subsequent studies, which have expanded the scope to cover: other cell types; 3D jelly roll regions; more complex decomposition reactions (e.g., electrolyte breakdown and short circuit events); and other cathode chemistries such as Nickel Manganese Cobalt (NMC) and Lithium Iron Phosphate (LFP)\cite{Kim2007,Peng2016,Coman2017, Bugryniec2020,Kong2021}. 
As an alternative to the Hatchard-based models, parameter estimation for generic Arrhenius type reaction kinetic ODEs have been proposed and successfully utilized in the literature \cite{Shelkea2022,ping2017modelling,feng2018coupled, bhatnagar2025chemical}. Many of these models build upon the fundamental principles of the Hatchard-type model while abstracting the specific components of the reaction decomposition. This enables the development of parametrically simple models that are more closely matched to experimental data. \\

Parameter estimation for thermal runaway models comes with significant challenges in determining the unknown reaction heat and kinetic parameters governing Arrhenius-based kinetic ODEs. This process typically relies on calorimetry data to measure chemical heat release. Accelerating Rate Calorimetry (ARC) is a widely used method for studying battery thermal runaway behavior at the single-cell level. Alternatively, other types of calorimetry data, such as Differential Scanning Calorimetry (DSC), can also be employed. In the literature, common fitting methods include linear approximations of the governing equations, genetic/population-based algorithms, machine learning techniques, or combinations of these methods \cite{chen2021simplified,ping2017modelling,Ren2018,wang2021thermal,bhatnagar2025chemical,koenig2023accommodating}. 
Linearisation methods are the most widely used, particularly for fitting ARC data. These methods involve dividing the temperature range into stages and linearising the Arrhenius kinetic ODEs within each stage. Parameters are then estimated using least-squares fitting. For example, \citet{chen2021simplified} and \citet{ping2017modelling} demonstrated temperature staging and linearisation approaches for ARC data fitting for a two-stage and five-stage model, respectively. Similarly, \citet{Ren2018} and \citet{wang2021thermal} combined linear fitting using the Kissinger method \cite{kissinger1956variation} with genetic algorithms to estimate the parameters of thermal runaway models based on DSC data.
However, the approximations involved in linearisation often reduce the model’s fidelity and robustness, leading to inaccurate representation of the underlying data. \citet{bhatnagar2025chemical} provides a detailed explanation of the limitations and pitfalls of linearisation. While some inaccuracies can be mitigated through extensive manual tuning, this approach at best yields a reasonable approximation of the original data.\\

\citet{bhatnagar2025chemical} and \citet{koenig2023accommodating} utilised Chemical Reaction Neural Networks \cite{ji2021autonomous} to fit Arrhenius ODE models to ARC and DSC experimental data. These networks explicitly encode kinetic reactions within their architectures and employ gradient-based optimization to directly estimate the equation parameters. The resulting models can be integrated over time and provide accurate predictions of thermal runaway behavior. However, this approach sometimes requires careful tuning of hyperparameters, such as the learning rate and the number of training steps, to achieve an optimal solution.\\

Population methods, such as Particle Swarm Optimization (PSO) \cite{kennedy1995particle}, have been widely used in the literature to solve parameter fitting problems for several battery modeling applications. \citet{cheng2024identification} tested various population methods to fit the RC parameters of equivalent electric circuit (EEC) models to the voltage output of an Extended Hybrid Pulse Power Characterization (EHPPC) test, obtaining low error in predicted terminal voltage. \citet{rahman2016electrochemical} used PSO to fit key parameters of a 1-D electrochemical model of a lithium-ion battery, and \citet{tian2025physics} applied a two-population method to fit the parameters of a pseudo-2D electrochemical model\cite{doyle1993modeling}. These works demonstrate the utility of population methods in solving parameter-fitting problems for battery physics models. In particular, PSO is a commonly used population method due to having a small number of hyper-parameters, low implementational complexity and accelerated convergence. Hence, using PSO to fit thermal runaway Arrhenius ODEs is a natural extension. PSO (or population methods in general) has several properties that make it well-suited for searching the parameter space of Arrhenius ODEs. Due to the stiffness of the governing ODEs, for large regions of the search space the computed jacobians of the loss function (obtained from integrating the ODEs) may be ill-conditioned. PSO does not require gradient computations, making it robust for these scenarios. Furthermore, PSO is inherently parallelizable, as the loss computation - the most expensive step of an optimization iteration for this application -  is independent for each particle. \\

Despite this perceived suitability, using population methods to search the parameter space of an N-stage Arrhenius ODE system (described in Section \ref{sec:TR_ODE_sys}) in a single optimization (referred to as the brute-force approach in this work) is computationally expensive due to the curse of dimensionality. \citet{chen2015measuring} studied the curse of dimensionality on population algorithms like PSO and Differential Evolution \cite{price2006differential}, concluding that parameter selection guidelines for population methods in low dimensional space may not scale to higher-dimensional search spaces, leading to inadequate exploration of the high-dimensional search space. Alternately, one can use a very large number of particles and/or search iterations, but it becomes prohibitively expensive to use large population sizes given that loss computation involves integrating a system of stiff ODEs. Furthermore, the Arrhenius ODE system is highly sensitive to parameter variations, requiring an exhaustive search of local minimums to find the optimal solution. \\ 

This work proposes a novel layered fitting method based on PSO to obtain accurate fits cost-efficiently for Arrhenius equation-based thermal runaway models from experimental ARC data. The method adopts a divide-and-conquer approach to address the curse of dimensionality, breaking the high-dimensional optimization problem into a series of sequential low-dimensional problems. Thermal runaway models are fit to two distinct ARC datasets using both layered PSO and brute-force PSO. The improved accuracy of the fits and lower computational cost of using the layered PSO relative to the naive brute-force method is demonstrated. The obtained parameters from layered PSO are also tested in 3D simulations of ARC and oven tests, demonstrating the generalizability of the models for broader thermal runaway Finite Element (FEM) simulations.

\section{Methods and Models}

\subsection{Thermal Runaway ODE System}
\label{sec:TR_ODE_sys}
The mathematical model for thermal runaway is based on a lumped-parameter approach, utilising a series of Arrhenius-based reaction equations. This model captures the decomposition reactions occurring during self-heating and the thermal runaway process. Specifically, the decomposition at various stages of the heating process can be expressed as follows:
\begin{equation}
    \label{eqn:arrhenius}
    \frac{dc_{i}}{dt}= f_{i}(c_{i})A_{i}e^{\frac{-E_{a,i}}{k_{b} T}}, 
    \textit{       i=1,2...,N} 
\end{equation}

\begin{equation}
    \label{eqn:conc_form}
    f_{i}(c_{i})= c_{i}^{n_{i}}(1-c_{i})^{m_{i}},
\end{equation}

where \(c_{i}\) is the normalized reactant concentration, \(A_{i}\) is the frequency factor, \(E_{a,i}\) is the activation energy, \(k_{b}\) is the Boltzmann constant, T is the temperature, and N refers to the number of stages in the model. \(m_{i}\) and \(n_{i}\) are reaction orders from the rate law. The reaction orders determine whether the reaction is \(n^{th}\) order (when $m_{i}=0$) or an auto-catalytic type i.e. the reaction increases as the product is generated ($m_{i}>0$) \cite{brown1997prout,grandjacques2021thermal}.\\

The heat \(Q_{i}\) from the \(i^{ih}\) stage exothermic decomposition reaction can be calculated as

\begin{equation}
    \label{eqn:heat_arrhenius}
    \dot{Q_{i}}=h_{i}\frac{dc_{i}}{dt},
\end{equation}

where \(h_{i}\) is the reaction enthalpy. The temperature update of the cell due to the thermal runaway can then be computed using

\begin{equation}
    \label{eqn:arrhenius_temp_update}
    m_{cell}c_{p}\frac{dT}{dt}=\sum_{i=1}^{N}\dot{Q_{i}},
\end{equation}

where $m_{cell}$ and \(c_{p}\) represent the mass and specific heat of the cell respectively. A good model representing the thermal runaway process is achieved by obtaining the values of the following kinetic and reaction heat parameters,

  \begin{equation}
        \label{eqn:trainable_params}
   \boldsymbol{\theta_{i}}=\boldsymbol{[}A,E_{a},h,m,n\boldsymbol{]}_{i}, \forall \text{i=1,2..N},
  \end{equation}
 such that the temperature and heat rate profiles (shown in Figure \ref{fig:data_demo}) predicted by integrating the model closely match the experimental ARC data.

\subsection{Issues with Conventional Fitting Methods}
\label{subsec:conventional_fitting_methods}

\subsubsection{Linearized Arrhenius Kinetics}
\label{sec:linearization_desc}
\begin{figure}[h!]
\centering
\begin{subfigure}{0.48\textwidth}
    \includegraphics[width=\textwidth]{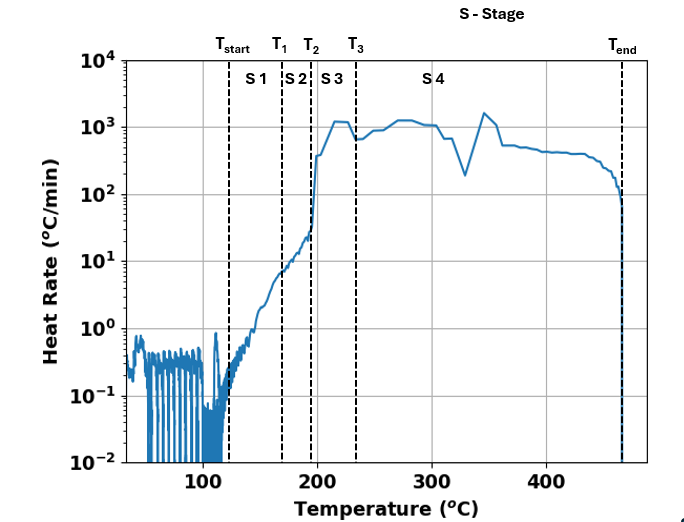}
    \caption{}
    \label{fig:div_4_stage}
\end{subfigure}
\hfill
\begin{subfigure}{0.49\textwidth}
    \includegraphics[width=\textwidth]{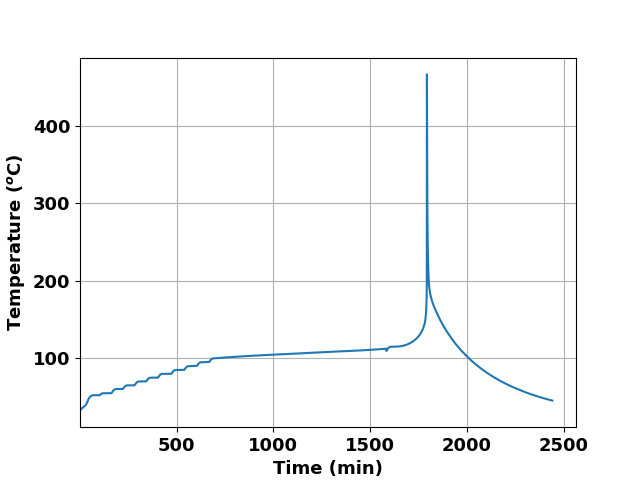}
    \caption{}
    \label{fig:time_temp}
\end{subfigure}
\hfill
\caption{\centering Data from an ARC test (\subref{fig:div_4_stage}) Heat rate data, with a sample division into N (=4) stages (\subref{fig:time_temp}) Temperature evolution with time }
\label{fig:data_demo}
\end{figure}

A widely used method for fitting the kinetic parameters of the Arrhenius ODE system, as described by Equation \ref{eqn:trainable_params}, involves linearizing the governing ODE system. The process begins by dividing the ARC data into stages, as shown in Figure \ref{fig:div_4_stage}. Each stage is represented by ODE in the form of equations (\ref{eqn:arrhenius}) and (\ref{eqn:heat_arrhenius}). For each stage, the ODE is fit independently for the self-heating data within a specific temperature range \([T_{i-1},T_{i}]\), with example boundary temperatures, \(T_{i}\), shown in Figure \ref{fig:div_4_stage}. The linear fit equations, derived through a series of linearizing assumptions described in \cite{chen2021simplified,sun2023thermal}, are given by

\begin{equation}
    \label{eqn:linearized_fit}
    ln \left[ \frac{dT}{dt} \right] =ln \bm{[} A_{i}(T^{end}_{i}-T^{start}_{i}) \bm{]} - \frac{E_{a,i}}{RT},
\end{equation}

\begin{equation}
    \label{eqn:enth_appx}
    h_{i} = \eta_{i} mc_{p}(T^{end}_{i}-T^{start}_{i}).
\end{equation}

where \(\eta_{i}\) is an adjustment factor for the enthalpy \cite{chen2021simplified}, obtained via manual tuning. While this method is straightforward, it suffers from several drawbacks. The linearization assumptions often do not hold true in practice and the re-using of certain parameters from previous studies reduces the accuracy and generalization of the fits. The pitfalls of the method have been discussed in detail by \citet{bhatnagar2025chemical}. 

\subsubsection{Particle Swarm Optimization}
To address the limitations highlighted in the previous section, a fitting method that can fit the governing ODEs in their original form (i.e. without linearisation) is sought. One effective approach for achieving this is a population-based optimization method such as Particle Swarm Optimization (PSO) \cite{kennedy1995particle}. PSO directly searches for the optimal parameters of the ODE system without relying on linearization assumptions. It is an iterative, zero-order algorithm that employs a population of \textit{particles} to explore the search space, with each particle updating its \textit{position} and \textit{velocity} based on the objective function. In this work, the optimization problem is framed as

\begin{equation}
  \min_{\textbf{u}} \; f(\textbf{u}),
\end{equation}

\begin{equation}
     u_{j}^{min} \leq u_{j} \leq u_{j}^{max} \;\;\;\; j=1,..,P,
\end{equation}

where  \(f:\mathbb{R^{P}} \mapsto \mathbb{R} \) is the objective function, \textbf{u} is the parameter vector, and P is the dimensionality of the search space. In the most general case, P=5N, where N is the number of stages. Every component of the search space is bounded by search limits, given by \([u_{j}^{min},u_{j}^{max}], \forall j=1,...,P\). During iterative updates of the population, the position \((\textbf{u}^{i}\)) and velocity \((\textbf{v}^{i}\)) of the \(i^{th}\) particle are updated using the following equations:

\begin{equation}
    \textbf{u}^{i}=\textbf{u}^{i}+\textbf{v}^{i},
\end{equation}

\begin{equation}
    \textbf{v}^{i}=w\textbf{v}^{i}+c_{1}r_{1}(\textbf{u}^{i}_{best}-\textbf{u}^{i})+ c_{2}r_{2}(\textbf{u}_{best}-\textbf{u}^{i}),
\end{equation}

where \(\textbf{u}_{best}^{i}\) represents the personal best position visited by a particle and \(\textbf{u}_{best}\) is the global best position visited by any particle of the swarm.  The parameters \(r_{1}\), \(r_{2}\)  are usually randomly chosen in each iteration between 0 and 1, while \(c_{1}\), \(c_{2}\), and \(w\) are predetermined iteration-dependent values. The initial and final values of these can be tuned to make the swarm more explorative (i.e explore regions of the permissible search space) or more exploitative (i.e identify the best minimizer within a local region) at any given iteration. \\

In this work, the value of \(h_{i}\) is approximated using Equation \ref{eqn:enth_appx}, with \(\eta_{i}\) being fit for each stage. The search space then takes the form

\begin{equation}
    \label{eqn:overall_search_space}
    \Theta_{all}= \bigcup_{i=1}^{N}\boldsymbol{\theta_{i}} =\bigcup_{i=1}^{N} [A,E_{a},\eta,m,n]_{i},
\end{equation}

 with \(\Theta_{all}\) representing the union of the sets of parameters for each stage ODE. The loss function takes the form:

\begin{equation}
    \label{eqn:loss_function}
    Loss= \lambda_{1} \sum_{t} \left ( log_{10}\left (\frac{dT}{dt}\right)_{data}- log_{10}\left(\frac{dT}{dt}\right)_{predicted} \right)^{2}+\lambda_{2} \sum_{t} (T_{data}-T_{predicted})^{2},
\end{equation}

 where \(\lambda_{1}\) and \(\lambda_{2}\) represent weighting coefficients for the loss terms, and \(\frac{dT}{dt}\) and T represent the heat rate and temperature respectively. The loss of the rate plot is computed on a log scale to account for the large variability in values observed. The search space limits for each parameter are described in Table \ref{tab:search_limits}. Reflective boundary conditions are applied at the search space limits in order to keep the particles within the defined search space.\\

 \begin{table}[h!]
    \centering
    \
\begin{tabular}{ccc} \toprule
    Parameter  & Lower Limit & Upper Limit \\ \midrule
    
     A\((s^{-1})\) & 1E8 & 1E25   \\ 
     \(E_{a}\)(J) & 1E-19 & 3.5E-19 \\
     \(\eta\) & 0.5 & 1.7 \\
     m & 0 & 8 \\
     n & 0 & 8 \\ \bottomrule
\end{tabular}
\caption{Search ranges for the Arrhenius ODE fitting problem}
    \label{tab:search_limits}
\end{table}

However, as previously discussed, using PSO in a brute-force manner to search the entire kinetic parameter space simultaneously is expensive and leads to inaccurate fits. This issue worsens with a larger value of N, which corresponds to a larger-dimensional search space, which requires a larger number of search particles.

\subsection{Layered Optimization Fitting Algorithm}
\label{subsec:layered_PSO}
\begin{figure}[h!]
\centering
\begin{subfigure}{0.6\textwidth}
    \includegraphics[width=\textwidth]{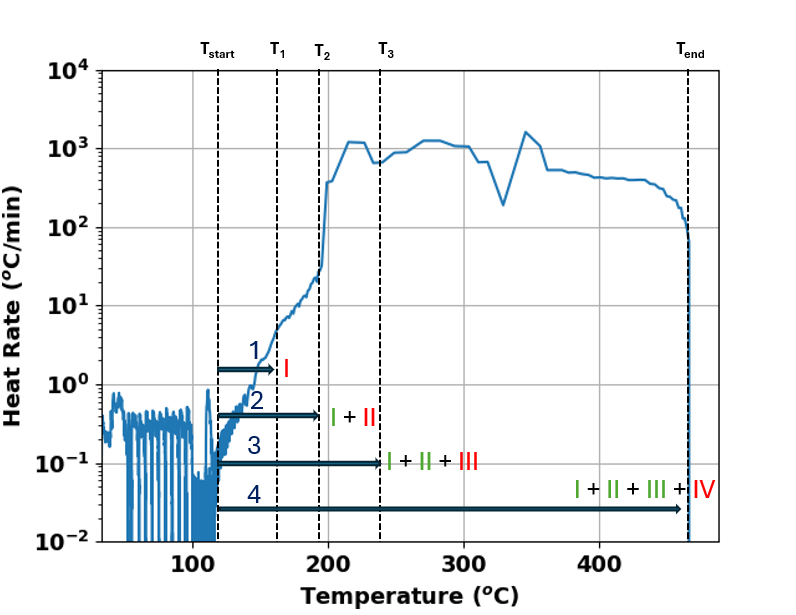}
    \caption{}
    \label{fig:layer_optimization}
\end{subfigure}
\hfill

\caption{\centering (\subref{fig:layer_optimization}) Visual depiction of the layered optimization algorithm. Each layer is shown by the dark blue arrow and an integer \(n\) and represents an optimization problem solved using PSO. The red Roman numeral indicates the stage fit in the \(n^{th}\) layer, and green Roman numerals depict previously fit stages used in the \(n^{th}\) layer. The heat contribution of each stage is summed in a given layer and compared to experimental data.}
\label{fig:layered_optimization_main}
\end{figure}

\begin{algorithm}
\setstretch{1.35}
    \caption{Layered Optimization Loss Computation Pseudocode for a Given Particle}\label{algo:layered_fitting}
    \begin{algorithmic}
        \REQUIRE $n_{layer},T_{start}, \boldsymbol{T}_{1,2,..n_{layer}},\left(\frac{dT}{dt}\right)_{data}, T_{data},\boldsymbol{\Theta}_{1,...,i_{layer}-1},\boldsymbol{\theta}_{particle}$

                \STATE $\left(\frac{dT}{dt}\right)_{predicted}=0$

                \FOR{$i_{stage}=1$ to $n_{layer}$}

                    \IF{$i_{stage}< n_{layer}$}
                        \STATE $ \boldsymbol{\theta} = \boldsymbol{\Theta}_{1,...,i_{layer}-1}[i_{stage}] $   \COMMENT{Use parameters fit in previous layer }
                    \ELSE

                        \STATE $ \boldsymbol{\theta}= \boldsymbol{\theta}_{particle}$ \COMMENT{Use parameters currently being optimized by PSO in layer \(n_{layer}\)}
                        
                    \ENDIF
                    
                    %\STATE $\left(\frac{dT}{dt}\right)_{predicted}(t)+=$ $integrate(T_{start},T_{n_{layer}},\boldsymbol{\theta}$) \COMMENT{Integrate ODE in \([T_{start},T_{i_{layer}}]\)}

                    \STATE $\dot{Q}_{i_{stage}} (t)=h_{i}(\boldsymbol{\theta})\frac{dc_{i}}{dt}(t), c_{i}(t)=\int_{T_{start}}^{T_{i_{stage}}} \frac{dc_{i}}{dt}(\boldsymbol{\theta},T) dT$ \COMMENT{Integrate ODE in \([T_{start},T_{i_{stage}}]\)}

                    \STATE $\left(\frac{dT}{dt}\right)_{predicted}(t)+=$ $\frac{1}{m_{cell}c_{p}}\dot{Q}_{i_{stage}} (t) $
                    
                \ENDFOR
                %\STATE $T_{predicted}(t)=integrate(\left(\frac{dT}{dt}\right)_{predicted}(t))$ \COMMENT{Integrate for Temperature History}

                \STATE $T_{predicted}(t)= \int_{t_{start}}^{t_{end}}\left(\frac{dT}{dt}\right)_{predicted}(t) dt$ \COMMENT{Integrate for Temperature History}
                
                \STATE $Loss= \lambda_{1} \sum_{t} \left( log_{10}\left (\frac{dT}{dt}\right)_{data}-log_{10}\left(\frac{dT}{dt}\right)_{predicted} \right)^{2}+\lambda_{2} \sum_{t} (T_{data}-T_{predicted})^{2}$
                
        \RETURN Loss
    \end{algorithmic}
    
\end{algorithm}

As outlined above, both the linear fitting and brute-force PSO encounter difficulties when fitting the parameters of Arrhenius ODE systems (Equation \ref{eqn:overall_search_space}). To address these difficulties, this work introduces a layered optimization algorithm based on PSO. The algorithm is designed to fit N-stage Arrhenius ODEs to ARC data in a computationally efficient manner. In particular,  the staging process used for the linear fitting method is combined with PSO to decompose a high-dimensional search into N low-dimensional searches, leading to a much more manageable problem. \\

Figure \ref{fig:layered_optimization_main} illustrates the layered fitting in visual form. Each layer is depicted by a dark blue arrow and is indexed by an integer \(n\). Similar to the linear fit, the heat rate plot is divided into N stages, with each stage defined by the start and end temperature. Each stage is assigned an ODE of the form of Equation \ref{eqn:arrhenius}. 
In the \(n^{th}\) layer, only the parameters corresponding to stage \(n\) (depicted by red Roman numerals in Figure \ref{fig:layered_optimization_main}) are fitted using PSO, while the parameters of previously fit stages (depicted by green Roman numerals in Figure \ref{fig:layered_optimization_main}) are kept constant. The heat rate contribution for each stage \(m=1,2,..,n\) is computed by integrating its corresponding ODE in the range \([T_{start},T_{n}]\) and summed to compute the total heat rate and temperature evolution over time. These outputs are compared to the raw data to compute the loss (Equation \ref{eqn:loss_function}). Algorithm \ref{algo:layered_fitting} outlines the pseudocode for the loss calculation of a single particle in a specific optimization layer \(n_{layer}\) and search iteration. \\

When fitting stage N (where N is the number of stages), the heat contribution of the last stage is not added to the overall heat rate until the temperature \(T_{N-1}\) is reached, i.e, \(\eta_{N}=0 \text{ if } T < T_{N-1}\). This is a modeling assumption aimed at capturing the effects of sudden, large heat release. \citet{sun2023thermal} proposed a similar model which used a fixed, pre-computed term as the heat released during the last stage, attributing the source to extensive internal short-circuiting. \citet{wang2021thermal} and \citet{feng2018coupled} used a step-function for the value of the frequency factor (\(A_{i}\)) of the anode, with the discontinuity temperature matching the separator collapse temperature as reported by \citet{feng2014thermal}. \\

The proposed layered PSO algorithm breaks down the optimization problem of maximum dimensionality 5N (solved by brute-force PSO) to N optimization problems of maximum dimensionality 5, as the fitting space for a single stage (Equation \ref{eqn:trainable_params}) has five parameters. An objective of this work is to demonstrate the improved accuracy of the layered PSO versus brute-force PSO, when the brute-force methods costs at least as much as the layered method. The relative cost for both methods can be tuned by setting the number of search particles accordingly.\\

Assume that the cost of integration per equation, per particle, for \(N_{ts}\) steps is C. The layered PSO algorithm involves a total of \(\sum_{i=1}^{N}i\) stiff ODE integrations per particle, where N is the number of stages. Hence, the cost of layered PSO fit in a single search iteration is 

\begin{equation}
\label{eqn:layered_PSO_cost}
    C_{layered}= C*n_{p,layered}*N*(N+1)/2
\end{equation}

where \(n_{p,layered}\) is the number of particles in the layered PSO fit. Similarly, in the brute-force method all N equations are fit simultaneously, and the cost of a single search iteration for brute-force PSO 

\begin{equation}
\label{eqn:brute_force_PSO_cost}
    C_{bf}= C*n_{p,bf}*N
\end{equation}

The cost of the brute-force fit is at least as much as the layered fit if

\begin{equation}
    n_{p,bf} \geq \frac{(N+1)}{2}*n_{p,layered}
\end{equation}

In this work, \(n_{p,layered}\) is set to 1,000, and \(n_{p,bf}\) is set to 10,000. Using a four stage model, this represents a 4x larger compute cost for the brute-force PSO. Despite the higher compute cost, it will be shown in the results that the brute-force PSO returns a less accurate fit compared to the layered PSO. Both algorithms are run for 50 search iterations each, with identical search space limits, PSO hyper-parameters, and particle position initialization schemes. All optimizations were run on a 12th Gen Intel(R) Core(TM) i7-12800H processor.

\subsection{Four stage thermal runaway model}
\label{sec:4_stage}
The four-stage thermal runaway model takes the form shown in equations (\ref{eqn:4_stage_model_1})-(\ref{eqn:enth_appx_second}) below: 

\begin{equation}
        \label{eqn:4_stage_model_1}
        \frac{dc_{1}}{dt}=c_{1}A_{1}e^{(\frac{-E_{a,1}}{RT})},
    \end{equation}

\begin{equation}
    \label{eqn:4_stage_model_2}
    \frac{dc_{2}}{dt}=c_{2}A_{2}e^{(\frac{-E_{a,2}}{RT})},
\end{equation}

\begin{equation}
    \label{eqn:4_stage_model_3}
    \frac{dc_{3}}{dt}=c_{3}^{n_{3}}(1-c_{3})^{m_{3}}A_{3}e^{(\frac{-E_{a,3}}{RT})},
\end{equation}

\begin{equation}
    \label{eqn:4_stage_model_4}
    \frac{dc_{4}}{dt}=c_{4}^{n4}(1-c_{4})^{m_{4}}A_{4}e^{(\frac{-E_{a,4}}{RT})},
\end{equation}

\begin{equation}
    \label{eqn:4_stage_model_energy}
    mc_{p}\frac{dT}{dt}= \sum_{i=1}^{4} h_{i} \frac{dc_{i}}{dt},
\end{equation}

 \begin{equation}
    \label{eqn:enth_appx_second}
    h_{i} = \eta_{i} mc_{p}(T^{end}_{i}-T^{start}_{i}).
\end{equation}

The first two stages are modeled as first-order (\(n=1,m=0\)) and the third and fourth stages approximated as auto-catalytic type. This simplification reduces the cost of fitting and generalizes the model equations used in \cite{chen2021simplified,Coman2017,ping2017modelling}, while retaining some previously used dynamics for individual stages that were found to work well. 

\subsection{3D Thermal Runaway Simulations}
 After fitting was done in 0D, 3D simulations were performed in the multi-purpose CFD software Altair\textsuperscript{\textregistered} AcuSolve\textsuperscript{\textregistered} using the fitted parameters. The reader is referred to Section \ref{sec:3D_details} for more details on the 3D cell model. To model thermal runaway in a computationally efficient manner, a dual time-marching algorithm was utilized. In particular, the system of Arrhenius equations was adaptively controlled using a sub-stepping algorithm within the primary time-marching scheme of the Energy equation. During the sub-stepping routine, the 0D Arrhenius ODE model was solved, and the nodal temperature values were updated. The method is equivalent to adding an additional source term \( S_{TR} \) to the energy equation e.g. 

\begin{equation}
    \label{eqn:3D_PDE}
    \rho c_{p}\frac{\partial T}{\partial t}= \nabla\cdot(\kappa\nabla T) + S_{TR},
\end{equation}

where is $\rho$ the density, $c_p$ is the heat capacity, $\kappa$ is the thermal conductivity, and the source term \( S_{TR} \) is given by

\begin{equation}
    \label{eqn:STR}
    S_{TR}=\sum_{i=1}^{N}{h_i\frac{dc_i}{dt}},
\end{equation}

where \(h_{i}\) represents the enthalpy of different cell components. $S_{TR}$ is active only within the jellyroll of the cell and is modeled using the four-stage thermal runaway model. In the ARC test, the cell undergoes adiabatic self-heating and no ambient temperature is required. For the oven test, the temperature reference is a single value representing the oven temperature. 
 
\section{Results and Discussion}

The layered and brute-force PSO fitting methods are demonstrated on two ARC datasets. Both datasets were for a Molicel 21700 P45B in an EV-ARC calorimeter, see \citet{giuliano2025experimental} for full details.  In the first test, the cell is placed in a holder and vented gases are allowed to be released; this test will be referred to as \textit{open} hereafter. In the second test, the cell is enclosed in a sealed canister within the calorimeter chamber, all gases are contained within the canister; this test will be referred to as \textit{closed} hereafter. These two tests represent a range of real-world thermal runaway scenarios, such as vented and sealed battery pack enclosures. Due to the sealed nature of the closed test, the temperatures recorded during thermal runaway were significantly higher.\\

The experiments focus on the four-stage model introduced in Section \ref{sec:4_stage}. In each case, it is shown that the layered PSO returns a more accurate fit to experimental ARC data than the brute-force method, despite the brute-force method taking longer times and being run with ten times more particles. The obtained model's generalizability is demonstrated by conducting 3D ARC and oven test simulations of the cylindrical cell.

\subsection{Open Test}
\label{sec:open_dataset_results}

 Table \ref{tab:staging_temperatures_4_stage_open} details the values of the temperatures at which the stages are divided, corresponding to Figure (\ref{fig:div_4_stage}). Note that these temperatures are empirical assumptions (as in the linearized fitting), and the values of these temperatures affect the fitting dynamics and enthalpies of the model. However, by fitting \(\theta_{i}, \forall i=1,..,N\), the proposed layered optimization algorithm reduces the dependence of the resulting model on these initially chosen temperature values, increasing the generalization capability of the model and resulting in significant improvement over the linear fitting, where \(\theta_{i}\) has to be manually tuned.  \\

\begin{figure}[h!]
\centering
\begin{subfigure}{0.45\textwidth}
    \includegraphics[width=\textwidth]{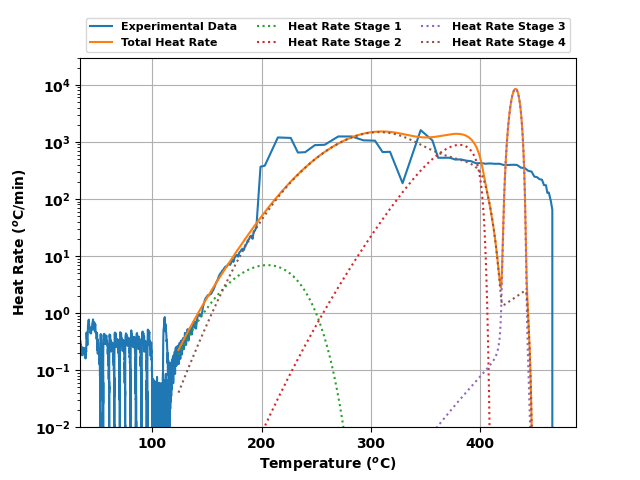}
    \caption{}
    \label{fig:rate_open_data_BF}
\end{subfigure}
\begin{subfigure}{0.45\textwidth}
    \includegraphics[width=\textwidth]{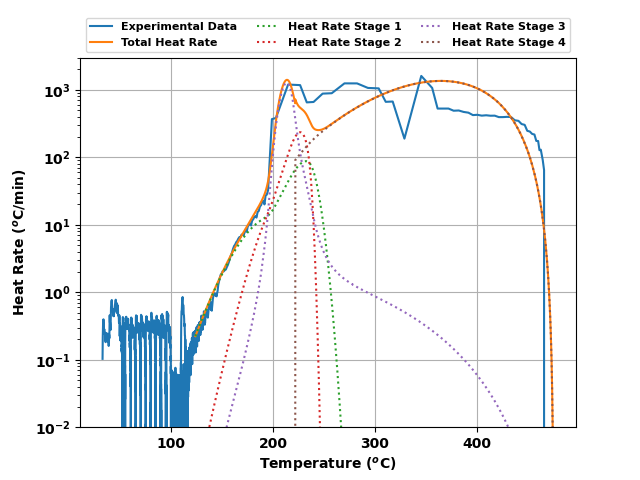}
    \caption{}
    \label{fig:rate_open_data_layered}
\end{subfigure}
\begin{subfigure}{0.45\textwidth}
    \includegraphics[width=\textwidth]{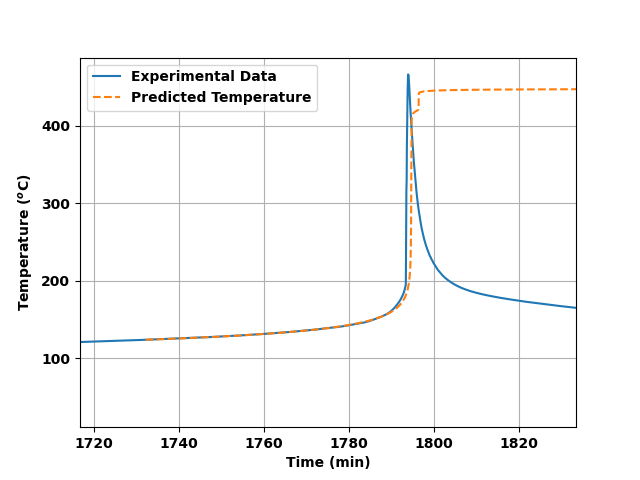}
    \caption{}
    \label{fig:temp_open_data_BF}
\end{subfigure}
\begin{subfigure}{0.45\textwidth}
    \includegraphics[width=\textwidth]{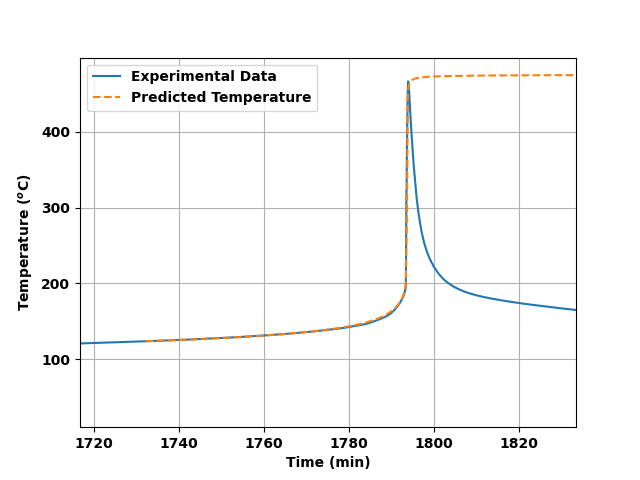}
    \caption{}
    \label{fig:temp_open_data_layered}
\end{subfigure}
\hfill

\caption{\centering Results from fitting data from the open test (\subref{fig:rate_open_data_BF}) Heat rate fit using brute-force PSO (\subref{fig:rate_open_data_layered})  Heat rate fit using layered PSO (\subref{fig:temp_open_data_BF}) Temperature prediction using brute-force PSO (\subref{fig:temp_open_data_layered}) Temperature prediction using layered PSO }
\label{fig:open_data_results}
\end{figure}

 Figure \ref{fig:rate_open_data_BF} shows the heat-rate fit from the brute-force method. The transition to faster self-heating around 200 \(^{o}C\) is not well captured, and the high heat-rate region is predicted to have unphysical oscillations including a predicted maximum heat rate that is much higher than experimental data. In contrast, the predicted heat rate plot using layered PSO (Figure \ref{fig:rate_open_data_layered}) shows a good match to experimental data, with the model predicting the dynamics of the thermal runaway well throughout the process. The initial slow temperature rise and the transition to faster self-heating are both well-captured. As per the modeling assumption, stage 4 does not contribute to the total heat until reaching the temperature \(T_{3}\) (shown in Table \ref{tab:staging_temperatures_4_stage_open}), after which stage 4 becomes the dominant source of heat release during the thermal runaway process. Figure \ref{fig:temp_open_data_BF} shows the predicted temperature from the brute-force fit, where the thermal runaway time and the peak temperature are not accurately predicted. In contrast, Figure \ref{fig:temp_open_data_layered} shows the predicted temperature from the layered PSO fit, accurately predicting both the key metrics. The difference in temperature prediction post thermal runaway is because cooling is not included in the thermal runaway simulation. This can be addressed by switching from an adiabatic environment to one that includes cooling post thermal runaway.\\
 
 The time taken for fitting using the brute force method was recorded at 3897 seconds, whereas the layered PSO took 1924 seconds to complete. Hence, despite the brute-force PSO having a higher compute cost, it returns a poorer fit to the ARC data.\\

\begin{figure}[h!]
\centering
\begin{subfigure}{0.45\textwidth}
    \includegraphics[width=\textwidth]{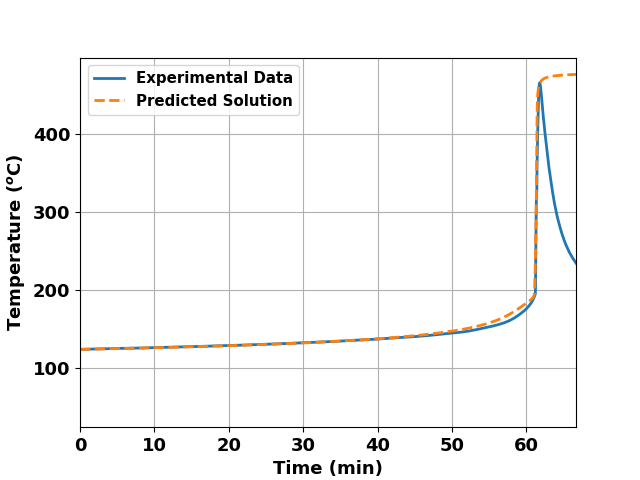}
    \caption{}
    \label{fig:temp_open_data_3D}
\end{subfigure}
\begin{subfigure}{0.45\textwidth}
    \includegraphics[width=\textwidth]{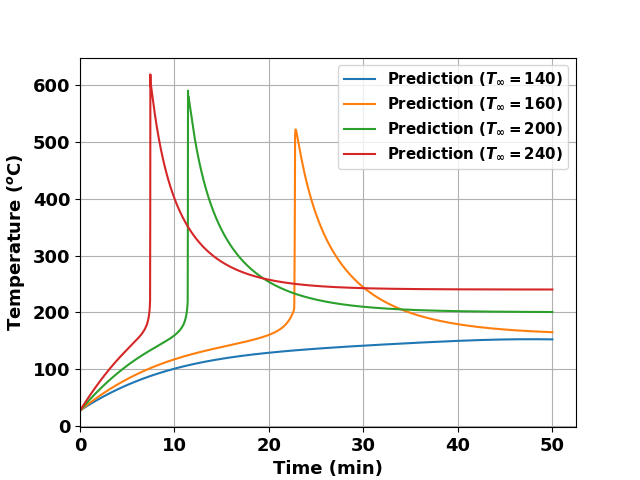}
    \caption{}
    \label{fig:oven_test_open}
\end{subfigure}
\hfill

\caption{\centering (\subref{fig:temp_open_data_3D}) 3D  simulation of self-heating and temperature comparison with experiment (\subref{fig:oven_test_open}) 3D oven test simulations at several temperatures}
\label{fig:open_data_results}
\end{figure}

The obtained parameters from the layered PSO, fit in 0D, are then used to simulate an ARC test in 3D. Figure \ref{fig:temp_open_data_3D} compares the predicted mean cell temperature during 3D simulation with the experimental data. Due to differences in the thermal dynamics of the cell between the 0D and 3D models (e.g. anisotropic transient conduction), the predicted thermal runaway time - assumed as the time when the temperature crosses 180 \(^{o}C\) - was approximately two minutes earlier than than the experimental time. Correcting this required a small adjustment (a 1.5 \% increase) in the value of the parameter \(E_{a,2}\). After this adjustment, the maximum temperature and the time of thermal runaway is predicted accurately in the 3D simulation using the calibrated model. Finally, oven tests are simulated at several temperatures using the obtained parameters. Figure \ref{fig:oven_test_open} shows the expected trend of thermal runaway for different oven temperatures. If the oven temperature is too low (\(T_{\infty}=140\text{ } ^o \)C), the cell does not go into thermal runaway. However, as the oven temperature increases, thermal runaway is observed, with a shorter time to thermal runaway and higher maximum temperature.

\begin{table}[h!]
    \centering
    \
\begin{tabular}{ccccc} \toprule
    \(T_{start}\) & \(T_{1}\) & \(T_{2}\) & \(T_{3}\)  \\ \midrule
    
     123 & 161 & 191 & 221  \\ \bottomrule
\end{tabular}
\caption{Staging temperatures for the open test, all values in \(^o\)C}
    \label{tab:staging_temperatures_4_stage_open}
\end{table}

\begin{table}[]
    \centering
    \
\begin{tabular}{ccccccc} \toprule
    \textbf{Stage No.}  & \textbf{I.C} \((c_{i,0}\)) & \(\boldsymbol{A_{i}} (s^{-})\)  &  \(\boldsymbol{E_{a,i}}\) (J) & \(\boldsymbol{h_{i}}\) (J) & \(\boldsymbol{m_{i}}\) & \(\boldsymbol{n_{i}}\) \\ \midrule
    
    \textbf{Stage 1}  & 1.0 & 3.23E15 & 2.495E-19 & 2894 & 0 & 1  \\ 
                             \bottomrule
     \textbf{Stage 2}  & 1.0 & 3.11E21 & 3.55E-19*       & 2285 & 0 & 1  \\ 
                \bottomrule

    \textbf{Stage 3}  & 0.04 & 2.59E24 & 3.50E-19       & 1345 & 3.00 & 3.14  \\ 
                \bottomrule

     \textbf{Stage 4}  & 0.04 & 1.00E8 & 1.56E-19    & \makecell{0 \((T < T_{3})\) \\ 18224 \((T \geq T_{3})\)} & 0 & 3.14  \\ 
                 \bottomrule
    
\end{tabular}
\caption{\centering Obtained parameters from the open test using layered PSO. The * indicates the parameter adjusted to account for 3D effects when running ARC and oven test simulations.}
    \label{tab:open_dataset_results_table}
\end{table}

\subsection{Closed Test}
\label{sec:closed_dataset_results}

\begin{figure}[h!]
\centering
\begin{subfigure}{0.45\textwidth}
    \includegraphics[width=\textwidth]{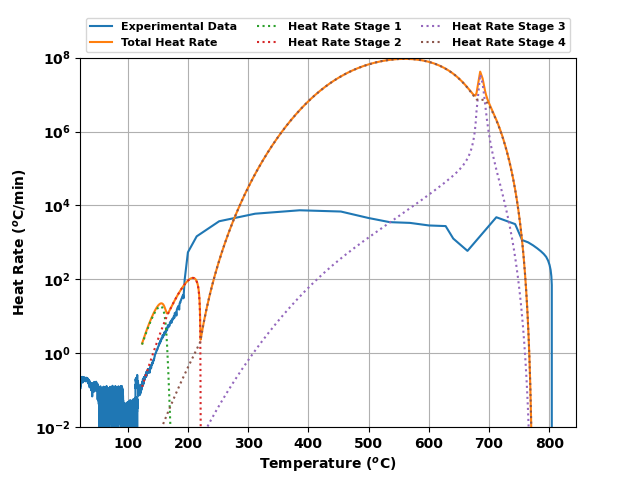}
    \caption{}
    \label{fig:rate_closed_data_BF}
\end{subfigure}
\hfill
\begin{subfigure}{0.45\textwidth}
    \includegraphics[width=\textwidth]{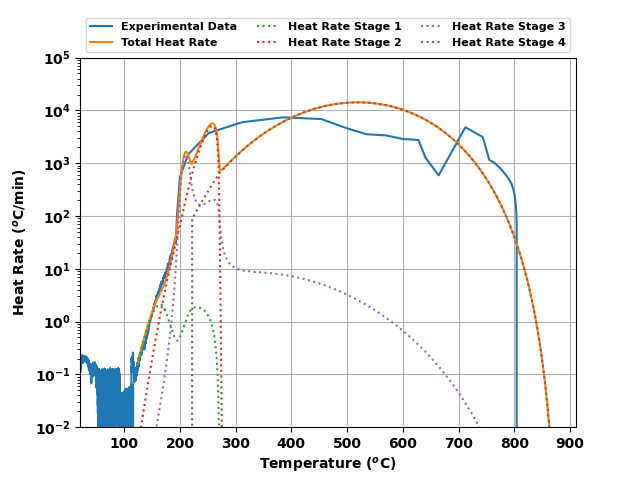}
    \caption{}
    \label{fig:rate_closed_data_layered}
\end{subfigure}
\hfill
\begin{subfigure}{0.45\textwidth}
    \includegraphics[width=\textwidth]{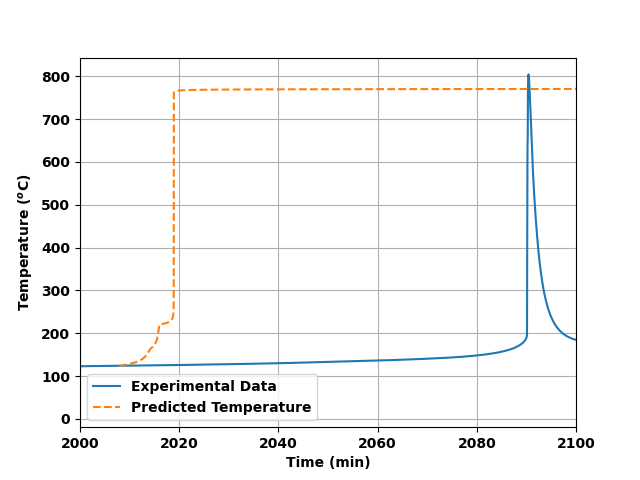}
    \caption{}
    \label{fig:temp_closed_data_BF}
\end{subfigure}
\hfill 
\begin{subfigure}{0.45\textwidth}
    \includegraphics[width=\textwidth]{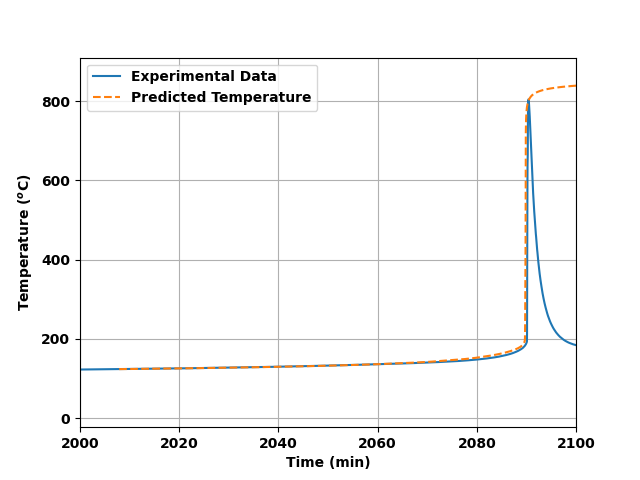}
    \caption{}
    \label{fig:temp_closed_data_layered}
\end{subfigure}
\hfill
\caption{\centering Results from fitting the data from the closed test (\subref{fig:rate_closed_data_BF})  Heat rate fit from brute-force PSO  (\subref{fig:rate_closed_data_layered}) Heat rate fit from layered PSO  (\subref{fig:temp_closed_data_BF}) Temperature prediction from brute-force PSO (\subref{fig:temp_closed_data_layered}) Temperature prediction from layered PSO}
\label{fig:closed_data_results}
\end{figure}

Table \ref{tab:staging_temperatures_4_stage_closed} details the temperatures of stage division for the closed test. Figure \ref{fig:closed_data_results} shows the results of the layered optimization fitting for the data obtained from the closed test. Figure \ref{fig:rate_closed_data_BF} shows the heat-rate prediction from the brute-force PSO fit. The model is unable to accurately capture the slow self-heating phase of thermal runaway, and greatly overpredicts the maximum heating rate during thermal runaway. Consequently, Figure \ref{fig:temp_closed_data_BF} shows that the thermal runaway is predicted far earlier than the experimental data, with the predicted peak temperature being underpredicted as well. Figure \ref{fig:rate_closed_data_layered} shows the fit from layered PSO, showing good agreement with experimental data. The initial slow temperature rise and the subsequent transition to rapid heat release are both well captured. Figure \ref{fig:temp_closed_data_layered} shows the predicted temperature, showing that the time of thermal runaway and maximum temperature are well predicted. Compared to the open test, this dataset has a higher self-heating rate and maximum temperature during thermal runaway. The layered fitting captures these differences effectively, yielding an excellent predictive model.  The obtained parameters from layered PSO are shown in Table \ref{tab:closed_dataset_results_table}. The time taken for fitting using the brute force method was recorded at 3346 seconds, whereas the layered PSO took 1998 seconds to complete. Once again, despite the brute-force PSO having a higher compute cost, it returns a poorer fit to the ARC data.\\

\begin{figure}[h!]
\centering
\begin{subfigure}{0.45\textwidth}
    \includegraphics[width=\textwidth]{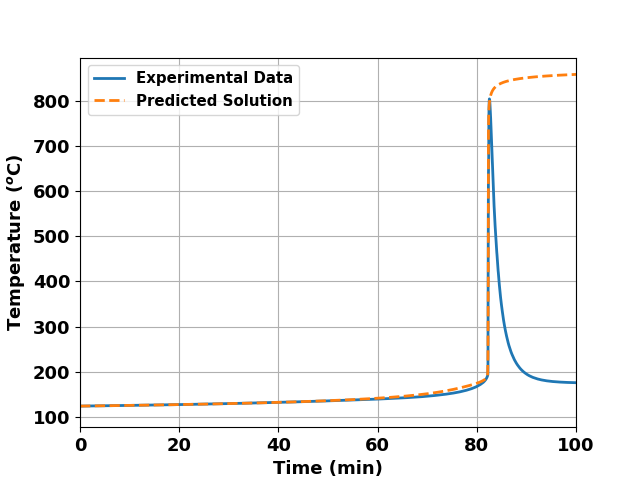}
    \caption{}
    \label{fig:temp_closed_data}
\end{subfigure}
\hfill
\begin{subfigure}{0.45\textwidth}
    \includegraphics[width=\textwidth]{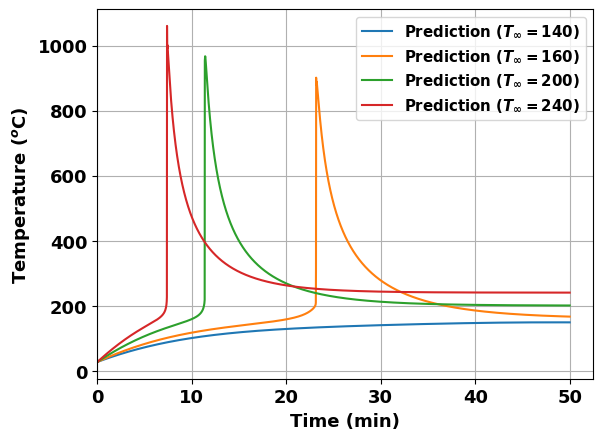}
    \caption{}
    \label{fig:oven_test_closed}
\end{subfigure}
\hfill
\caption{\centering  (\subref{fig:temp_closed_data}) 3D simulation of self-heating and temperature comparison with experimental data (\subref{fig:oven_test_closed}) 3D oven test simulations at several temperatures}

\end{figure}

The parameters were subsequently included in a 3D ARC simulation, as previously described. As with the open test, due to differences in the thermal behaviour of the 0D and 3D models, the predicted thermal runaway time in the 3D simulation occurred approximately 4 minutes early. By increasing the parameter \(E_{a,2}\) by 0.75\%, the 3D simulation shows a good agreement with the experimental data, as shown in Figure \ref{fig:temp_closed_data}, predicting accurately the thermal runaway time and the maximum temperature. Finally, oven tests were undertaken for several temperatures, as shown in Figure \ref{fig:oven_test_closed}. The results demonstrate the expected trend: higher oven temperatures lead to quicker thermal runaway and higher maximum temperatures. It may be noted that the maximum temperatures achieved in the oven tests from the closed test are much higher than those from the open test for the same oven temperatures. These observations align with the expected dynamics of the closed test, specifically larger self-heating rates and higher temperatures. As observed in the previous section, if the oven temperature is too low, the model predicts that thermal runaway does not occur.

\begin{table}[h!]
    \centering
    \
\begin{tabular}{cccc} \toprule
    \(T_{start}\) & \(T_{1}\) & \(T_{2}\) & \(T_{3}\)  \\ \midrule
    
     123 & 151 & 201 & 221  \\ \bottomrule
\end{tabular}
\caption{Staging temperatures for the closed test, all values in \(^o\)C}
    \label{tab:staging_temperatures_4_stage_closed}
\end{table}
 
\begin{table}[]
    \centering
    \
\begin{tabular}{ccccccc} \toprule
    \textbf{Stage No.}  & \textbf{I.C} \((c_{i,0}\)) & \(\boldsymbol{A_{i}} (s^{-})\)  &  \(\boldsymbol{E_{a,i}}\) (J) & \(\boldsymbol{h_{i}}\) (J) & \(\boldsymbol{m_{i}}\) & \(\boldsymbol{n_{i}}\) \\ \midrule
    
    \textbf{Stage 1}  & 1.0 & 1.261E17 & 2.697E-19 & 2133 & 0 & 1  \\ 
                             \bottomrule
     \textbf{Stage 2}  & 1.0 & 3.96E21 & 3.525E-19*       & 3809 & 0 & 1  \\ 
                \bottomrule

    \textbf{Stage 3}  & 0.04 & 1.00E25 & 3.500E-19       & 1448 & 3.40 & 4.23  \\ 
                \bottomrule

     \textbf{Stage 4}  & 0.04 & 1.00E8 & 1.556E-19    & \makecell{0 \((T < T_{3})\) \\  43994 \((T \geq T_{3})\)} & 0 & 6.56  \\ 
                 \bottomrule
    
\end{tabular}
\caption{\centering Obtained parameters for the closed test. The * indicates the parameter adjusted to account for 3D effects when running ARC and oven test simulations. }
    \label{tab:closed_dataset_results_table}
\end{table}

\subsection{Discussion}

The previous sections demonstrated the flexibility and accuracy of the layered PSO method in fitting N-stage Arrhenius ODE models to ARC data, in particular relative to brute-force PSO. The layered optimization technique can capture the differing dynamics of different ARC datasets efficiently and accurately, and the obtained models can generalize well to realistic 3D simulations of thermal runaway. Moreover, the method is agnostic to cell stoichiometry and geometry, and can generalize to an arbitrary number of stages. The method is not limited to PSO and the user can use any population method within the algorithm.\\

The proposed method allows users to fit models to ARC data with greater generality than previously possible, avoiding the need to reuse or adapt parameters from other literature \cite{Feng2016,feng2018coupled,ostanek2020simulating}, use linearization techniques ( Section \ref{sec:linearization_desc}) and avoiding the excessive computational costs associated with brute-force fitting using population methods like PSO. The method allows straightforward fitting of Arrhenius kinetics to the self-heating and thermal runaway stages of an ARC test, addressing limitations of other methods, such as directly computing the self-heating rate from the ARC data or using cumulative enthalpy methods. These alternative methods often do not generalize well outside the adiabatic environment of an ARC test simulation. Some common issues include inaccurate prediction of the maximum temperature when the heat input is increased and and difficulties with propagation. Recently, the use of Chemical Reaction Neural Networks (CRNNs) \cite{bhatnagar2025chemical} has also been shown to fit Arrhenius kinetics without making limiting assumptions on the model. However, a drawback of the CRNN method is that it requires a good initial condition for the parameters to be fit. This can be partly achieved using the linearization method, but may require further tuning of the initial condition in some cases to achieve a good fit. The layered PSO method searches the entire parameter space while iterating, and does not require a good initial condition.\\

Several recent studies have explored using machine learning models to predict the onset of thermal runaway without using Arrhenius ODE models. \citet{kim2023modeling} used Physics Informed Neural Networks (PINNs)\cite{raissi2019physics} to predict thermal runaway in lithium-ion batteries, successfully predicting thermal runaway for cases where labelled data was unavailable. \citet{jeong2024prediction} used physics-informed DeepONets \cite{goswami2023physics} to predict thermal runaway, embedding physical principles directly into the loss function. \citet{goswami2024advancing} used a coupled Graph Neural Network (GNN) and Long Short-Term Memory (LSTM) network to predict thermal runaway, leveraging virtual sensor data to predict thermal runaway during battery operation. However, these approaches rely on using data from simulations (e.g. finite element based simulations) to train the models, which still require an accurate model to fit the thermal runaway dynamics of the cell under investigation. This work hence enables and improves a broad range of predictive modeling approaches for thermal runaway, enabling safer battery pack designs and the creation of early warning systems for battery pack fire occurrence.

\section{Conclusion}

This work proposes a novel layered optimization fitting algorithm that incorporates a divide-and-conquer approach to fitting N-stage Arrhenius ODE models to Accelerating Rate Calorimetry (ARC) data. The proposed algorithm ensures the accuracy and generalizability of the fit models by eliminating model simplifications usually made to enable the fitting of the ODE system, while tackling the curse of dimensionality observed while optimizing in high-dimensional parameter spaces using population-based algorithms, by breaking the problem into several low dimensional optimization problems. The proposed algorithm is used to fit four-stage models and is found to return models that accurately reproduce the experimental data. The resultant models are simulated in 3D ARC and oven tests, the results of which show good agreement with experimental data and expected trends. The algorithm is demonstrated on two different ARC datasets, showing the flexibility of the proposed method. The proposed algorithm can also extend to several cell types, stoichiometries, and an arbitrary number of stages. Future work involves extending the method to other forms of calorimetry such as DSC, and fitting on ensembles of calorimetry datasets to enable uncertainty quantification of obtained parameters. This work represents a promising step forward in using simulation tools to design safe battery-powered systems efficiently and accurately.

\section{CRediT Authorship Statement}

\hspace{5mm} \textbf{Saakaar Bhatnagar}:
Conceptualization, Formal Analysis, Investigation, Methodology, Software, Visualization, Writing- Original Draft

\textbf{Andrew Comerford}: 
Conceptualization, Software, Formal Analysis, Project Administration, Writing- Original Draft

\textbf{Zelu Xu}: 
Conceptualization, Software, Formal Analysis, Validation, Writing- Review and Editing

\textbf{Simone Reitano}:  Data Curation, Investigation, Resources, Writing- Review and Editing

\textbf{Luigi Scrimieri}: Data Curation, Investigation, Resources, Writing- Review and Editing

\textbf{Luca Giuliano}: Data Curation, Validation, Writing- Review and Editing

\textbf{Araz Banaeizadeh}:  Project Administration, Supervision, Writing- Review and Editing

\section{Funding Sources}

This research received no specific grant from funding agencies in the public, commercial, or not-for-profit sectors.

\bibliographystyle{unsrtnat}
\bibliography{bib}

\appendix

\section{Appendix}

\subsection{Cell properties for 3D simulation}
\label{sec:3D_details}

\begin{table}[]
    \centering
    \
\begin{tabular}{cccc} \toprule
    Component & Density (\(kg/m^3\)) & Specific Heat (\(J/kgK\)) & Conductivity (\(W/mK\))  \\ \midrule
    
     \textbf{Jelly Roll} & 161 & 191 & (\(k_{r},k_{\theta},k_{h}\))=(0.3,28,28)  \\  \midrule

     \textbf{Steel Can} & 7917 & 460 & 14  \\ \midrule

     \textbf{Tabs} & 2770 & 986 & 175  \\ 
     \bottomrule
\end{tabular}
\caption{Properties of cell components simulated}
    \label{tab:3d_sim_physical_properties}
\end{table}

\begin{figure}[h!]
\centering
\begin{subfigure}{0.6\textwidth}
    \includegraphics[width=\textwidth]{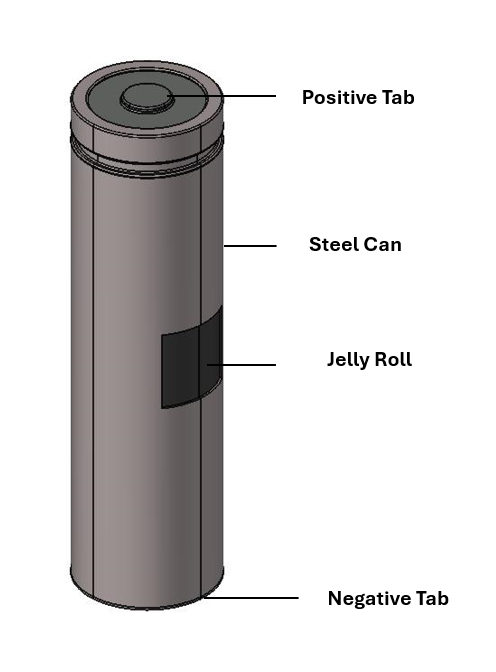}
    \caption{}
    \label{fig:cell_diagram}
\end{subfigure}
\hfill

\caption{\centering (\subref{fig:cell_diagram}) Geometry of the cell under consideration}
\label{fig:cell_diagram_main}
\end{figure}

 This section contains additional information about the 3D ARC and oven test simulations. The governing Equation (\ref{eqn:3D_PDE}) is discretized using a Galerkin least squares stabilized finite element method. The battery geometry consisted of three main components: the jellyroll; the aluminium positive and negative tabs/terminals; and a steel can. A summary of the material properties for each of these components is provided in Table \ref{tab:3d_sim_physical_properties}. The jellyroll, an effective material representation of the true wound geometry, has cylindrical anisotropic properties for thermal conductivity; the through-plane conductivity is significantly lower than in-plane conductivity. Figure \ref{fig:cell_diagram} shows the 3D model of the cell used.

\end{document}